# A Workflow for GLAM Metadata Crosswalk

Arianna Moretti[1], Ivan Heibi[2], Silvio Peroni[3].
[1] University of Bologna, Italy – arianna.moretti4@unibo.it
[2] University of Bologna, Italy – ivan.heibi2@unibo.it
[3] University of Bologna, Italy – silvio.peroni@unibo.it

## ABSTRACT

The acquisition of physical artifacts not only involves transferring existing information into the digital ecosystem but also generates information as a process itself, underscoring the importance of meticulous management of FAIR data and metadata. In addition, the diversity of objects within the cultural heritage domain is reflected in a multitude of descriptive models. The digitization process expands the opportunities for exchange and joint utilization, granted that the descriptive schemas are made interoperable in advance. To achieve this goal, we propose a replicable workflow for metadata schema crosswalks that facilitates the preservation and accessibility of cultural heritage in the digital ecosystem. This work presents a methodology for metadata generation and management in the case study of the digital twin of the temporary exhibition "The Other Renaissance - Ulisse Aldrovandi and the Wonders of the World". The workflow delineates a systematic, step-by-step transformation of tabular data into RDF format, to enhance Linked Open Data. The methodology adopts the RDF Mapping Language (RML) technology for converting data to RDF with a human contribution involvement. This last aspect entails an interaction between digital humanists and domain experts through surveys leading to the abstraction and reformulation of domain-specific knowledge, to be exploited in the process of formalizing and converting information.

## 1. INTRODUCTION

The digitization process offers the advantage of extending access to cultural heritage geographically and chronologically [24, 25]. Additionally, to foster preservation in the digital ecosystem, the acquisition of artifacts must be accompanied by FAIR metadata creation, management, and maintenance [18, 29]. Temporary exhibitions introduce further elements of interest and complexity in formalizing GLAM (Galleries, Libraries, Archives, and Museums) data [16], as they often feature heterogeneous objects from multiple sources, potentially described according to different representation models. Thus, conversions between formats and *schema crosswalks* between data models are needed to exploit data jointly for specific purposes.

The term *schema crosswalk* refers to the mapping between conceptualization systems that describe at least partially overlapping domains, intending to identify points of contact and divergence to facilitate data exchange. In this context, mappings between schemas for metadata management in the description of digital objects have been created [23], sometimes accompanied by invitations to best practices, such as harmonizing existing schemas and using recommended data types to avoid semantic loss [8].

In this paper, we introduce a workflow for simplifying the creation and, thus, reproducibility of schema crosswalks exploiting digital humanists' and domain experts' contributions. This approach seeks to balance the reuse of general components with solutions developed ad hoc for the case study, i.e. the creation of the digital twin of the temporary exhibition "The Other Renaissance - Ulisse Aldrovandi and the Wonders of the World"[1] [3, 4]. Specifically, we focus on providing a process for converting data in tabular form, describing the objects included in the exhibition and the digitization process for creating their digital replicas, into the Resource Description Framework (RDF) format. The goal is to enable the use of Semantic Web technologies on the content of exhibition material descriptions and acquisition metadata, formulated as Linked Open Data (LOD).

---

[1] https://site.unibo.it/aldrovandi500/en/mostra-l-altro-rinascimento

The paper includes a Literature Review section, providing an overview of several approaches for exploiting structured data as RDF. The Case Study of the temporary exhibition "The Other Renaissance - Ulisse Aldrovandi and the Wonders of the World" is then introduced, to illustrate the context of application of the workflow. The Methodology section presents the procedure step-by-step, providing reasons for the adopted choices. In the Future Developments, we introduce some potential and desirable evolutions of the approach, to deliver a more complete and reusable research product. In the Conclusions, we summarise the content of the paper to highlight the key aspects of the proposed workflow.

## 2. LITERATURE REVIEW

The conditions to perform schema crosswalks are closely tied to a system's interoperability, meant as the property of a data model to exchange information seamlessly. On this topic, the EOSC Executive Board FAIR Working Group's Interoperability Task Force produced a report [7] outlining directives for facilitating data exchange, with a focus on the FAIR principle of interoperability across technological, semantic, organizational, legal, and syntactic levels. The document provides a crosswalk of data models, controlled vocabularies, and aggregators' guidelines, aiming to harmonize existing patterns for mapping and prevent semantic reductions due to metadata quality loss. Similarly, Milan Ojsteršek published a noteworthy crosswalk on Zenodo that maps the basic properties of widely used vocabularies [23], creating a comprehensive knowledge base for future developments in the metadata exchange domain.

A practical situation in which it is required to perform metadata crosswalks is the digital objects integration within collectors. Such activity often involves adapting data and metadata to meet the needed target formats and models, and users typically have to face challenges based on platform documentation clarity, familiarity with target formats, model knowledge, and technical tool proficiency. Indeed, many services share the common drawback of limited declarative support to the user for performing the crosswalk. Coherently, among the causes for metadata quality loss in the description of digital objects, the EOSC Interoperability Framework Report mentions: (1) mismatches between the updates of artifacts, vocabularies, metadata management software, and administered data; (2) compelled metadata conversion adjustments, and (3) unsupervised inclusion of freely structured metadata in the content aggregators [7]. Thus, a workflow for guiding the crosswalk process for uploading and updating data, and facilitating the selection of conceptual and technical tools for information exchange between formats, would be beneficial.

More in detail, in the Semantic Web domain, it is necessary to transform other formats into RDF or to query them as if they were RDF-structured. Some direct conversion tools that allow various formats to be accessed as if they were formalized in RDF enable such conversion but do not aim to create a unique abstraction for format management. Examples of this category are Any23[2], JSON2RDF[3], and CSV2RDF[4] [9].

Other approaches also offer an abstraction over the particular format used to define data. Such tools include D2RQ [6], a system for accessing the contents of relational databases as RDF graphs, thanks to a server in which a conversion of SPARQL query to SQL is implemented. On the other hand, Triplify [2] was among the first software components to be integrated into other applications for converting data from relational databases to RDF. In this technology, the semantic content is maintained by mapping HTTP-URI requests to SQL queries.

Other solutions include the RDF Mapping Language (RML) [12], a generic mapping language with customizable rules independent of specific implementations. RML was a forerunner in the reengineering approaches of formats according to the RDF model as a consequence of the use of SPARQL, also enabling the querying of relational databases. This technology was selected for the workflow step concerning the conversion to RDF for its accurate documentation and the wide set of open-source tools provided by the developers to facilitate the pipeline execution. In addition to that, the Aldrovandi temporary exhibition case study foresaw the necessity to convert information stored in CSV tables with non-trivial structures into RDF serializations. This circumstance motivated the adoption of RML also because of the possibility of extending the mapping rules through the definition of custom functions in Java and Turtle to potentially meet any specific need concerning the conversion process. A valuable alternative that was taken into account was Facade-x [8]: a generic meta-model that allows querying resources as if they were structured in RDF using wrappers, without extending the SPARQL syntax. It is concretely implemented in SPARQL Anything, a reengineering system that facilitates querying any structured data using SPARQL and creating knowledge graphs without requiring specific skills in a particular mapping language or an in-depth understanding of formats. However, the process implies automated reengineering to the target meta-model and domain knowledge reframing by an RDF and SPARQL expert. At a technical level, it implements a set of transformers mapped to various media types, extendable with Java classes. In addition to the above-mentioned tools, SPARQL Generate [19] should be mentioned as a declarative transformation language extending the SPARQL syntax for generating RDF graphs or textual streams, that can be further integrated with mediating query languages such as XPath[5] to handle new sources.

---

[2] https://any23.apache.org/
[3] https://github.com/AtomGraph/JSON2RDF
[4] https://clarkparsia.github.io/csv2rdf/

## 3. CASE STUDY

The temporary exhibition "The Other Renaissance - Ulisse Aldrovandi and the Wonders of the World", hosted at the Palazzo Poggi Museum in Bologna for six months starting from December 2022, showcased hundreds of artifacts belonging to the naturalist Ulisse Aldrovandi (1522-1605). To acknowledge its cultural significance, an experimental approach to save temporary exhibitions was undertaken [3].

In this context, a methodology for collecting, formalizing, managing, and exporting FAIR cultural heritage metadata was defined. More in detail, the specificity of this case study was closely related to the nature of the project of digitizing a temporary exhibition, envisaging the necessity to manage both the information derived from museum descriptions concerning the exhibited objects and the data generated during the digital acquisition process.

The metadata generation and management process was structured in steps. The first stage involved the tabular formalization of information on the museum objects and the acquisition processes, for the creation of two input datasets. Specifically, the first of these was structured based on metadata inferred from museum captions and - where possible - from catalogs' data. The result was a table with the following fields: Identification Number, Linked Identification Number, Relationship, Exhibition Room, Caption, Consistency, Documentary Typology, Technique, Reproduction Typology in Exhibition, Subjects, Original Title, Museum Title, English Title, Date, Discoverer, Author, Translator, Illustrator, Engraver, Publisher, Place of Publication, Museum Preparer, Commissioner, Parent Work Typology, Parent Work Title, Volume, Collection, Conservator Entity, Location of Conservation, Placement, Source, Digital Image, and Iconography. The table concerning the acquisition process, on the other hand, had a more complex structure, involving fields with various internal subdivisions. At the most general level, it provided data about Identification Numbers, Objects, Showcases, Captions, Current Status, Link, Notes, Acquisition, Processing, Modeling, Optimisation, Export, Metadata, and Upload.

In the meantime, to semantically describe the collected data, an Application Profile named CHAD-AP was conceptualized [5, 11] based on CIDOC CRM and CRM Digital, with a module for each of the two tables. Once the tables had been exported to CSV, they could be converted into RDF. From this point on, the traceability of the source information is maintained [21] by exploiting the OpenCitations data model [10]. Finally, the process ends with a quantitative analysis, visualization, and narration of the data using the tool MELODY [26].

This work focuses on the aspects related to converting the source dataset into RDF serializations. Our proposal (i.e., workflow) integrates software tools and human contributions from digital humanists and domain experts (in this case, museum curators).

## 4. METHODOLOGY

We present a workflow, summarised in Figure 1, for converting GLAM metadata collections available in tabular form to N-Triples RDF serializations, highlighting the human role in its design and execution. At a technical level, the procedure relies on LimeSurvey [20], parsers for JSON and YAML [28], and PYRML [22] for producing the output collection from the input dataset, exploiting RML mapping rules.

---

[5] https://www.w3.org/TR/xpath/

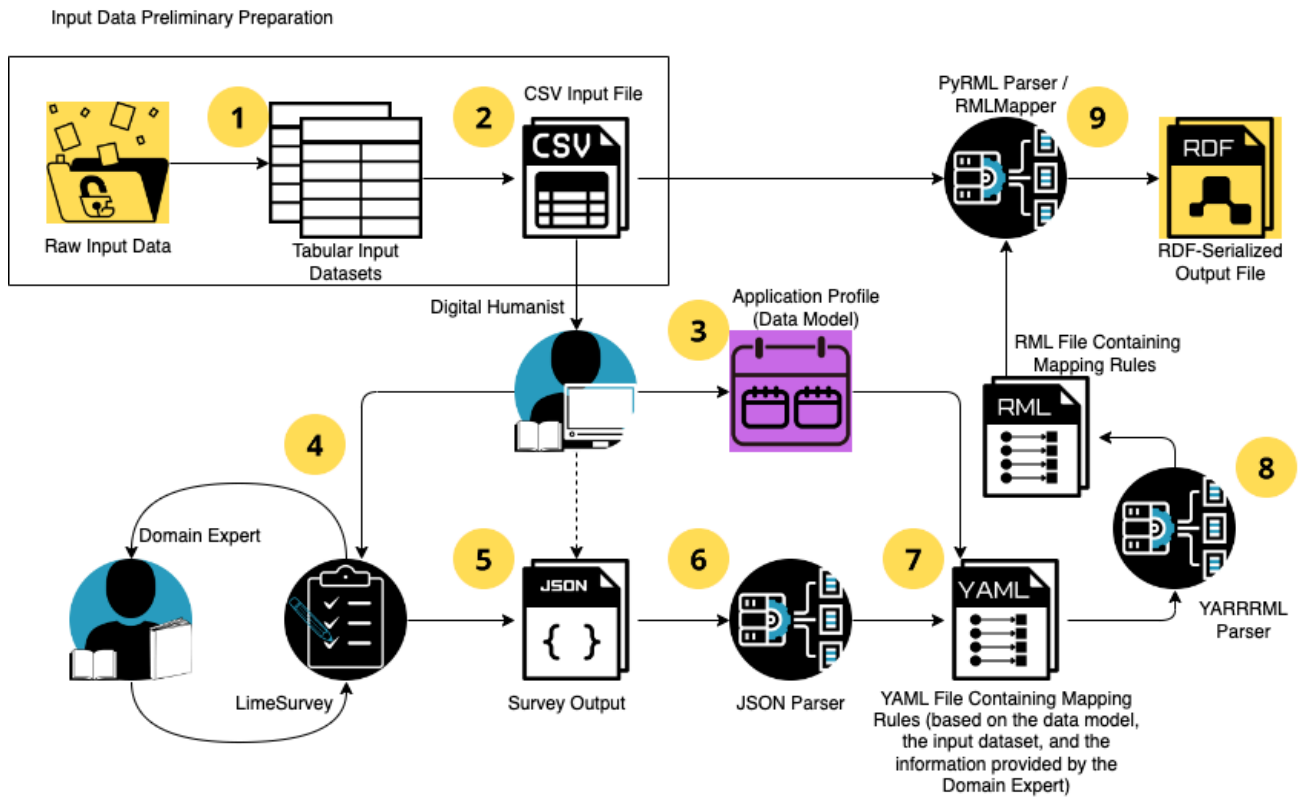

**Figure 1. A workflow for CSV to RDF schema crosswalk of GLAM data involving human contributions.**

Since the selection of the conversion technology contributes to determining the possibilities of querying and exploiting the generated data, as well as the structure of the workflow and the involvement of domain experts, the choice for the mediation technology has fallen on RML. In addition to the aforementioned customizability, among the advantages of this technology is the opportunity to express the basic domain knowledge in configuration files, without extending software or dealing with the complexities of direct use of technical tools.

The process follows a series of well-structured steps (see Figure 1). First of all, (1) the information obtained from the museum descriptions and the digital acquisition process is formalized in two separate tabular datasets, (2) that are then exported as CSV files. Unless an existing data model was reused to structure the input tables, during the first stage (3) the digital humanist examines the datasets and starts formulating an appropriate Application Profile, including one module for each of the datasets to describe, to exhaustively represent the gathered data. The obtained data model should have reached a stable version before starting the formulation of the RML rules. Afterward, (4) the digital humanist, with consolidated knowledge in the use of RML and associated technologies, formulates a LimeSurvey questionnaire to be presented to the domain expert, intending to guide formalizing their knowledge. This step aims to verify that the input information was correctly structured in the tables (i.e., that the museum descriptions were seamlessly interpreted) and that the data model does allow to convey all the informative content of the datasets. In addition to that, the questionnaire is an opportunity to pose additional questions to the domain experts, both to possibly improve the data model through unstructured text feedback and to gather additional preferences on the conversion output, where no formal constraints are posed by the data model or customization is needed. For example, as for the most recent model update, CHAD-AP imposes a constraint on the use of controlled vocabularies. For what concerns the creation techniques, the model defines an Expression (*frbroo:F2_Expression*) as a reference to the intellectual content of the object, which is generated through a creation event (*frbroo:F28_Expression_Creation*), that uses various creation techniques (*crm:P32_used_general_technique*), whose domain is an Activity (*crm:E7_Activity*) and whose range is a Type (*crm:E55_Type*). The information to be transposed in RDF is in the "Technique" field of the CSV file of the object dataset and, in the case study, it was expressed as a textual string (e.g.: "drawing technique"). However, to make the RDF transposition fully compliant with the data model, we adopted the Art & Architecture Thesaurus[6] codes (e.g.: aat:300054196, "drawing (image-making)"), and thus ad hoc functions had to be defined for performing the conversion. Different case studies might involve specific modifications to the Application Profile to be efficiently reused and properly represent the input data. Therefore, a survey question on adopting the same convention enables either use or don't the ad hoc conversion function. The following steps involve (5) exporting the questionnaire responses in JSON format to allow (6) the digital humanist to analyze the output and decide whether to review the data model and adjust the JSON file, in case the survey revealed further or unexpected information. (7) The rest of the data, along with the data model, is then used as input for the JSON parser to compile a YAML configuration file compliant with the syntax specification of the human-readable text-based representation for declarative mapping generation rules, named YARRRML [15]. At this stage of the process, (8) the obtained YAML file is used as input for the YARRRML parser, converting the configuration file into an RML Turtle serialized file. Eventually, (9) the input dataset and the RML configuration file are taken in input by the PyRML parser, to finally produce the RDF output dataset. As an alternative, this step can be performed by using the official Java RMLMapper [27].

---

[6] https://www.getty.edu/research/tools/vocabularies/aat/

It is worth noting that while it is possible to bypass the JSON-YAML conversion step by directly generating the RML file, it is recommended to follow the complete procedure since this helps limit the risk of errors by maintaining a gradual approach and providing an information formalization in a human-readable format.

The human contribution is condensed in the interaction between the domain expert and the digital humanist through the questionnaire formulation and compilation. In addition, questionnaires have been used to further optimize the process of formalizing domain information from experts who may lack coding skills. After evaluating options such as Microsoft Forms, SoGoSurvey, Google Forms, Typeform, SurveyMonkey, and Qualtrics, the choice fell to LimeSurvey. This tool, developed for academic and institutional purposes, is open source, freely accessible online, offers a widely usable free plan, can be configured to comply with GDPR, and allows the export of responses in various structured formats [17], a fundamental aspect for the subsequent conversion into RML configuration files. Further, with the perspective of adopting the same Application Profile to convert to RDF similarly structured collections, it ensures both the reuse of survey sections and the drafting of new custom-made questions, thus enabling the application of conversion patterns common for most metadata crosswalks and the expression of new requests to grasp the understanding of case-specific aspects. It is noteworthy that, while most of the information can be automatically formalized through a priori formulation of questions, new elements may emerge from the domain expert answers, causing the occasional intervention of the digital humanist on the data model. Furthermore, this human-centered step also implies verification of the JSON output's structural correctness before proceeding, in case of inconsistencies.

## 5. FUTURE DEVELOPMENTS

The presented case study allowed the definition of a reusable workflow that simplifies the creation of metadata schema crosswalk of GLAM data, from tabular to RDF format. However, considering the opportunities for joint use of variably structured cultural heritage data, a future development goal is to extend the workflow, making it potentially reusable for conversions between any-to-any formats. As an ambitious project, an intermediate milestone is proposed to achieve coverage in both input and output of widely used formats, including JSON, RDF, tabular, and XML, with their respective serializations.

Another potential evolution involves transitioning from a linear to an iterative workflow structure, where the human intervention limited to an initial phase is replaced by a "Human in the Loop" approach [1]. In this perspective, the domain expert is consulted again at the end of the process and participates in the evaluation of the output, guided in the formalization of their opinion through another questionnaire. In this way, the feedback received could be reintegrated as an additional input into the process, aiming to produce incremental versions of the output data until an optimal result is achieved.

Finally, the current workflow involves the direct and concatenated use of several software components in various programming languages. We argue that the reproducibility of the methodology could be enhanced by leveraging open-access tools for formalizing the procedure by providing examples of the execution phases and orchestrating the functioning of the technologies involved. In the first place, the steps of the procedure can be structured in an actionable workflow released on dedicated platforms that guarantee an adequate versioning system and free access to content, such as the Social Sciences & Humanities Open Marketplace[7] and Protocols[8]. The reproducibility potential could be further enhanced by tools that automatically execute the software components in a specific order, managing the inputs and outputs of intermediate steps. In this regard, MITAO [13][14] represents a valid choice, ensuring declarative management automation and process scalability. The software provides a user-friendly visual interface that facilitates users without programming skills to integrate data and tools in a customizable visual workflow and share the defined product. Additionally, the software can be extended and customized with the help of code developers (e.g., for integrating a specific tool for exporting LimeSurvey questionnaire results in JSON or converting it to YAML). As for the drawbacks, the current limitation of MITAO lies in its language-specific setup and it should, therefore, be extended to allow the combination of Python components with those in other programming languages, such as Java.

---

[7] https://marketplace.sshopencloud.eu/
[8] https://www.protocols.io/

## 6. CONCLUSIONS

We presented a workflow for facilitating cultural heritage metadata conversions between different formats and models, fostering collaborative use, and leveraging Semantic Web technologies. The RML-based methodology was introduced alongside the Aldrovandi temporal exhibition case study on which it was tested and refined, with particular attention to balancing the automation of replicable steps and the necessity of case-specific ad hoc customizations.

In particular, we outlined the value of human contribution in the context of data management in a workflow for the RDF formalization of the metadata collections concerning the temporary exhibition's objects and their digital acquisition process. Significant emphasis was placed on the interaction between digital humanists and domain experts and on the formalization of knowledge in machine-readable formats through custom online questionnaires. Potential integrations and adaptations have been outlined, such as extending the range of supported formats, adopting a "Human in the Loop" approach, and introducing a potential solution for an automated, executable, and reproducible workflow.

## 7. ACKNOWLEDGMENTS

This work was partially funded by Project PE 0000020 CHANGES - CUP B53C22003780006, NRP Mission 4 Component 2 Investment 1.3, funded by the European Union - NextGenerationEU.

## REFERENCES

[1] Anderson, Marc, and Karën Fort. "Human Where? A New Scale Defining Human Involvement in Technology Communities from an Ethical Standpoint." *The International Review of Information Ethics* 31, no. 1 (November 3, 2022). https://doi.org/10.29173/irie477.

[2] Auer, Sören, Sebastian Dietzold, Jens Lehmann, Sebastian Hellmann, and David Aumueller. "Triplify: Light-Weight Linked Data Publication from Relational Databases." In *Proceedings of the 18th International Conference on World Wide Web - WWW '09*, 621. Madrid, Spain: ACM Press, 2009. https://doi.org/10.1145/1526709.1526793.

[3] Balzani, Roberto, Sebastian Barzaghi, Gabriele Bitelli, Federica Bonifazi, Alice Bordignon, Luca Cipriani, Simona Colitti, et al. "Saving Temporary Exhibitions in Virtual Environments: The Digital Renaissance of Ulisse Aldrovandi – Acquisition and Digitisation of Cultural Heritage Objects." *Digital Applications in Archaeology and Cultural Heritage* 32 (March 1, 2024): e00309. https://doi.org/10.1016/j.daach.2023.e00309.

[4] Barzaghi, Sebastian, Federica Collina, Francesca Fabbri, Federica Giacomini, Alice Bordignon, Roberto Balzani, Gabriele Bitelli, et al. "Digitisation of Temporary Exhibitions: The Aldrovandi Case." In *Eurographics Workshop on Graphics and Cultural Heritage*, 181–83. The Eurographics Association, 2023. https://doi.org/10.2312/gch.20231176.

[5] Barzaghi, Sebastian, Ivan Heibi, Arianna Moretti, and Silvio Peroni. "Developing Application Profiles for Enhancing Data and Workflows in Cultural Heritage Digitisation Processes." arXiv, April 18, 2024. https://doi.org/10.48550/arXiv.2404.12069.

[6] Bizer, Christian, and Andy Seaborne. "D2RQ—Treating Non-RDF Databases as Virtual RDF Graphs." *World Wide Web Internet and Web Information Systems*, January 1, 2005.

[7] Corcho, Oscar, Magnus Eriksson, Krzysztof Kurowski, and Milan Ojsteršek. *EOSC Interoperability Framework: Report from the EOSC Executive Board Working Groups FAIR and Architecture*. Univerza v Mariboru, Fakulteta za elektrotehniko, računalništvo in informatiko, 2021. https://www.afs.enea.it/project/madia/Documenti/web/docs/EOSC-Interoperability-Framework-KI0221055ENN.pdf.

[8] Daga, Enrico, Luigi Asprino, Paul Mulholland, and Aldo Gangemi. "Facade-X: An Opinionated Approach to SPARQL Anything." In *Studies on the Semantic Web*, edited by Mehwish Alam, Paul Groth, Victor de Boer, Tassilo Pellegrini, Harshvardhan J. Pandit, Elena Montiel, Víctor Rodríguez Doncel, Barbara McGillivray, and Albert Meroño-Peñuela. IOS Press, 2021. https://doi.org/10.3233/SSW210035.

[9] Daga, Enrico, Luca Panziera, and Carlos Pedrinaci. "BASIL: A Cloud Platform for Sharing and Reusing SPARQL Queries as Web APIs," January 1, 2015.

[10] Daquino, Marilena, Silvio Peroni, David Shotton, and Arcangelo Massari. "The OpenCitations Data Model," 2020. https://doi.org/10.6084/M9.FIGSHARE.3443876.V7.

[11] /DH.arc. "Dharc-Org/Chad-Ap." Jupyter Notebook. 2024. Reprint, /DH.arc, April 17, 2024. https://github.com/dharc-org/chad-ap.

[12] Dimou, Anastasia, Miel Vander Sande, Pieter Colpaert, Ruben Verborgh, Erik Mannens, and Rik Van de Walle. "RML: A Generic Language for Integrated RDF Mappings of Heterogeneous Data." In *Proceedings of the Workshop on Linked Data on the Web Co-Located with the 23rd International World Wide Web Conference (WWW 2014), Seoul, Korea, April 8, 2014*, edited by Christian Bizer, Tom Heath, Sören Auer, and Tim Berners-Lee, Vol. 1184. CEUR Workshop Proceedings. CEUR-WS.org, 2014. https://ceur-ws.org/Vol-1184/ldow2014\_paper\_01.pdf.

[13] Ferri, Paolo, Ivan Heibi, Luca Pareschi, and Silvio Peroni. "MITAO: A User Friendly and Modular Software for Topic Modelling." *puntOorg International Journal* 5, no. 2 (October 24, 2020): 135–49. https://doi.org/10.19245/25.05.pij.5.2.3.

[14] Heibi, Ivan, Silvio Peroni, Luca Pareschi, and Paolo Ferri. "MITAO: A Tool for Enabling Scholars in the Humanities to Use Topic Modelling in Their Studies." In *AIUCD 2021 - Book of Extended Abstracts*, 175–82, 2021. https://doi.org/10.6092/unibo/amsacta/6712.

[15] Heyvaert, Pieter, Ben De Meester, Anastasia Dimou, and Ruben Verborgh. "Declarative Rules for Linked Data Generation at Your Fingertips!" In *The Semantic Web: ESWC 2018 Satellite Events*, edited by Aldo Gangemi, Anna Lisa Gentile, Andrea Giovanni Nuzzolese, Sebastian Rudolph, Maria Maleshkova, Heiko Paulheim, Jeff Z. Pan, and Mehwish Alam, 213–17. Lecture Notes in Computer Science. Cham: Springer International Publishing, 2018. https://doi.org/10.1007/978-3-319-98192-5_40.

[16] Kapsalis, Effie. *The Impact of Open Access on Galleries, Libraries, Museums, & Archives*, 2016.

[17] Klieve, Helen, Wendi Beamish, Fiona Bryer, Robyn Rebollo, Heidi Perrett, and Jeroen Van Den Muyzenberg. "Accessing Practitioner Expertise Through Online Survey Tool LimeSurvey." Griffith Institute for Educational Research, 2010. http://hdl.handle.net/10072/36611.

[18] Koster, Lukas, and Saskia Woutersen-Windhouwer. "FAIR Principles for Library, Archive and Museum Collections: A Proposal for Standards for Reusable Collections." *The Code4Lib Journal*, no. 40 (May 4, 2018). https://journal.code4lib.org/articles/13427.

[19] Lefrançois, Maxime, Antoine Zimmermann, and Noorani Bakerally. "Flexible RDF Generation from RDF and Heterogeneous Data Sources with SPARQL-Generate." In *Knowledge Engineering and Knowledge Management*, edited by Paolo Ciancarini, Francesco Poggi, Matthew Horridge, Jun Zhao, Tudor Groza, Mari Carmen Suarez-Figueroa, Mathieu d'Aquin, and Valentina Presutti, 10180:131–35. Lecture Notes in Computer Science. Cham: Springer International Publishing, 2017. https://doi.org/10.1007/978-3-319-58694-6_16.

[20] LimeSurvey. "LimeSurvey," January 26, 2024. Software Heritage. https://archive.softwareheritage.org/swh:1:dir:c36ac88e2358050ad4d00ca457aef1ce6d0a3180;origin=https://github.com/LimeSurvey/LimeSurvey;visit=swh:1:snp:cf6c3f1c990cfc9f0f91e295eace369de35df0a3;anchor=swh:1:rev:a8b1f5d8b5b45d3e5a9eba4fbeb92ecd2398cdb2.

[21] Massari, Arcangelo, Silvio Peroni, Francesca Tomasi, and Ivan Heibi. "Representing Provenance and Track Changes of Cultural Heritage Metadata in RDF: A Survey of Existing Approaches," 2023. https://doi.org/10.48550/ARXIV.2305.08477.

[22] Nuzzolese, Andrea Giovanni, and Giulio Settanta. "pyRML," 2024. https://archive.softwareheritage.org/swh:1:dir:7c7daa4e9e44e40be9b69b90130ed2c675df0cf6;origin=https://github.com/anuzzolese/pyrml;visit=swh:1:snp:7434d2226b78603ce653fffadf575387e26419cd;anchor=swh:1:rev:cd42ca3a92297c3527b537e24f42202a0295a4a7.

[23] Ojsteršek. "Crosswalk of Most Used Metadata Schemes and Guidelines for Metadata Interoperability." Zenodo, January 5, 2021. https://doi.org/10.5281/ZENODO.4420115.

[24] Peinado-Santana, Sara, Patricia Hernández-Lamas, Jorge Bernabéu-Larena, Beatriz Cabau-Anchuelo, and José Antonio Martín-Caro. "Public Works Heritage 3D Model Digitisation, Optimisation and Dissemination with Free and Open-Source Software and Platforms and Low-Cost Tools." *Sustainability* 13, no. 23 (January 2021): 13020. https://doi.org/10.3390/su132313020.

[25] Pescarin, Sofia. "Museums and Virtual Museums in Europe: Reaching Expectations." *SCIRES-IT - SCIentific RESearch and Information Technology* 4, no. 1 (April 30, 2014): 131–40. https://doi.org/10.2423/i22394303v4n1p131.

[26] Renda, Giulia, Marilena Daquino, and Valentina Presutti. "Melody: A Platform for Linked Open Data Visualisation and Curated Storytelling." arXiv, June 26, 2023. http://arxiv.org/abs/2306.14832.

[27] RMLio. "RMLio/Rmlmapper-Java." Java. 2018. Reprint, RDF Mapping Language (RML), April 1, 2024. https://archive.softwareheritage.org/swh:1:dir:bc91cd466746b8e6ae65d183589501dfafeb5df2;origin=https://github.com/RMLio/rmlmapper-java;visit=swh:1:snp:c3deeafbc6826c89dac0b831992ba21f858e7874;anchor=swh:1:rev:f8d15d97efb9a30359b05f37a28328584fe62744.

[28] ———. "YARRRML Parser," November 23, 2023. Software Heritage. https://archive.softwareheritage.org/swh:1:dir:b2a545fb1ca9a8e42898db5413bdb2cd98af909b;origin=https://github.com/RMLio/yarrrml-parser;visit=swh:1:snp:9d7369e64633ab91c03d48938ea542cb38a81366;anchor=swh:1:rev:52e74c0bf4a554ac0d5e376378c50565803a67bb.

[29] Wilkinson, Mark D., Michel Dumontier, IJsbrand Jan Aalbersberg, Gabrielle Appleton, Myles Axton, Arie Baak, Niklas Blomberg, et al. "The FAIR Guiding Principles for Scientific Data Management and Stewardship." *Scientific Data* 3, no. 1 (March 15, 2016): 160018. https://doi.org/10.1038/sdata.2016.18.